\documentclass[a4paper]{jpconf}
\usepackage{graphicx}
\usepackage{amssymb}
\begin{document}
\title{A self-consistent study of magnetic field effects on hybrid stars}

\author{V. Dexheimer}
\address{Department of Physics, Kent State University, Kent OH 44242 USA}

\author{B. Franzon}
\address{FIAS, Ruth-Moufang 1, 60438 Frankfurt am Main, Germany}

\author{S. Schramm}
\address{FIAS, Ruth-Moufang 1, 60438 Frankfurt am Main, Germany}

\ead{vdexheim@kent.edu}

\begin{abstract}
It is understood that strong magnetic fields affect the structure of neutron stars. Nevertheless, many calculations for magnetized neutron stars are still being performed using symmetric solutions of Einstein's equations. In this conference proceeding, we review why this is not the correct procedure and we also discuss the effects of magnetic fields on the stellar population and temperature profiles.
\end{abstract}

\section{Introduction}
\hfill \break

In addition to being dense and, to some extent, hot, neutron stars rotate extremely fast and possess strong magnetic fields.  According to observations of star periods and period derivatives,
magnetars can have a surface magnetic field as large as $10^{14}-10^{15}$ G \cite{Melatos:1999ji} and one expects to find even stronger magnetic fields inside these stars. One calculation estimates the internal magnetic field of a star to be of the order of $10^{16}$ G \cite{Makishima:2014dua} but, according to virial theorem arguments, neutron stars can have central magnetic fields of the order of  $10^{18}$ G depending on their composition, among other factors \cite{fushiki1989surface,lai1991cold,Cardall:2000bs,Ferrer:2010wz}. Unfortunately, the origin of strong magnetic fields in compact stars is still unclear.

\section{Modelling}
\hfill \break

A self-consistent study of strong magnetic field effects on neutron stars must contain two parts:
\medskip
\begin{enumerate}
\item Effects of magnetic fields on the equation of state (EoS)
\item Effects of magnetic fields on the solutions of Einstein's equations
\end{enumerate}

To illustrate the first item, we make use of the non-linear realization of the SU(3) sigma model (also referred to as Chiral Mean Field or CMF model) \cite{Papazoglou:1998vr,Dexheimer:2008ax,Dexheimer:2015qha}. This is an effective quantum relativistic model within the mean field approximation that describes hadrons interacting via meson exchange. The model is constructed from symmetry relations, which allows it  to be chirally invariant at high densities and/or temperatures.

In addition, the model also contains quarks and an effective potential for the field related to deconfinement, which has non-zero values for any density/chemical potential and/or temperature as long as the quarks are present \cite{Dexheimer:2009hi,Hempel:2013tfa}. In this way, different kinds of phase transitions (from smooth crossovers to strong first order ones) can be reproduced and the interactions with the medium determine what are the degrees of freedom present at a certain temperature and baryon chemical potential. In the low temperature case, a first order phase transition is found and, due to the assumption of global charge neutrality, a mixed phase is reproduced.

At low densities, the CMF Model is constrained by nuclear physics data, while at high temperatures and densities it is fitted to lattice QCD results and astrophysical observations, respectively. As an additional test of the formalism, we show in Fig.~1 the EoS for neutron star matter and compare it to perturbative QCD (PQCD) results for 3-flavor quark-gluon plasma at zero temperature including beta-equilibrium and charge neutrality \cite{Fraga:2013qra}.

\begin{figure}[t!]
\includegraphics[trim={0 1.74cm 0 0},clip,width=22pc]{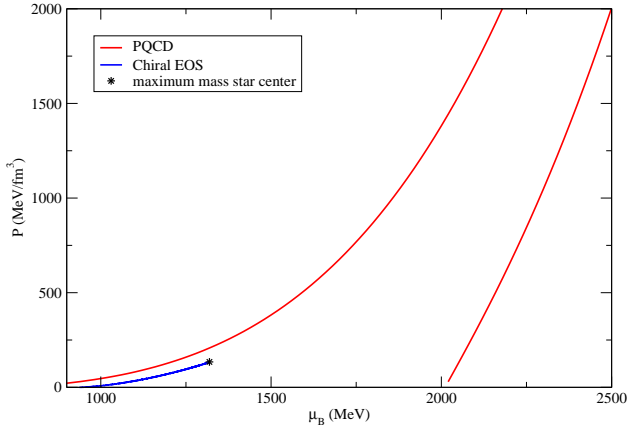}\hspace{2pc}%
\begin{minipage}[b]{14pc}
\caption{\label{label} CMF Model equation of state (pressure versus baryon chemical potential) for neutron stars in comparison with PQCD results. The error band reflects uncertainties due to renormalization. PQCD data from Ref~\cite{Fraga:2013qra}.\\ \\ \\}
\end{minipage}
\end{figure}

The effects of magnetic fields in the EoS are included in the CMF model though Landau quantization, the anomalous magnetic moment (AMM) for hadrons and the magnetization. The first one increases the amount of charged particles in the system \cite{Broderick:2000pe}. The third one modifies the pressure in the direction perpendicular to the magnetic field and creates an asymmetry in the EoS (independent of the pure magnetic field  asymmetry), which is larger at low temperatures (see Ref.~\cite{Strickland:2012vu} and references therein for details).

The anomalous magnetic moment induces a difference in the amount of spin up and spin down fermions in the star \cite{Broderick:2001qw}. Particles with spin projection opposite to their charge sign cannot have a zeroth Landau level, even when their orbital angular momentum is equal to zero, causing them to have
a smaller density when summing over all levels. Particles with spin projection equal
to their charge sign are enhanced. This causes, for example, the electrons to be fully polarized in the quark phase. For non-charged baryons, the ones with negative spin projection and negative AMM coupling are suppressed due to the increase in their effective masses. The baryons with negative spin projection and positive coupling are enhanced by the AMM. Particles with spin up have the exactly opposite behavior. For a figure showing the particle population for different spin projections see Ref.~\cite{Dexheimer:2012qk}.

For the solutions of Einstein's equations in the presence of strong magnetic fields, an axisymmetric configuration must be used. In addition, Maxwell's equations must be fulfilled in all parts of the star. To guarantee these, a macroscopic poloidal magnetic field is generated through currents in a numerical code which is part of the LORENE C++ class library for numerical relativity \cite{Bonazzola:1993zz,Bocquet:1995je}. In this formalism, the strength of the magnetic field throughout the star is controlled by either the dipole magnetic moment $\mu$ or the dimensionless current function $f_0$.

\section{Stellar Structure}
\hfill \break

\begin{figure}[t!]
\begin{minipage}{16pc}
\centering
\includegraphics[trim={0 0 0 1.75cm},clip,width=15pc]{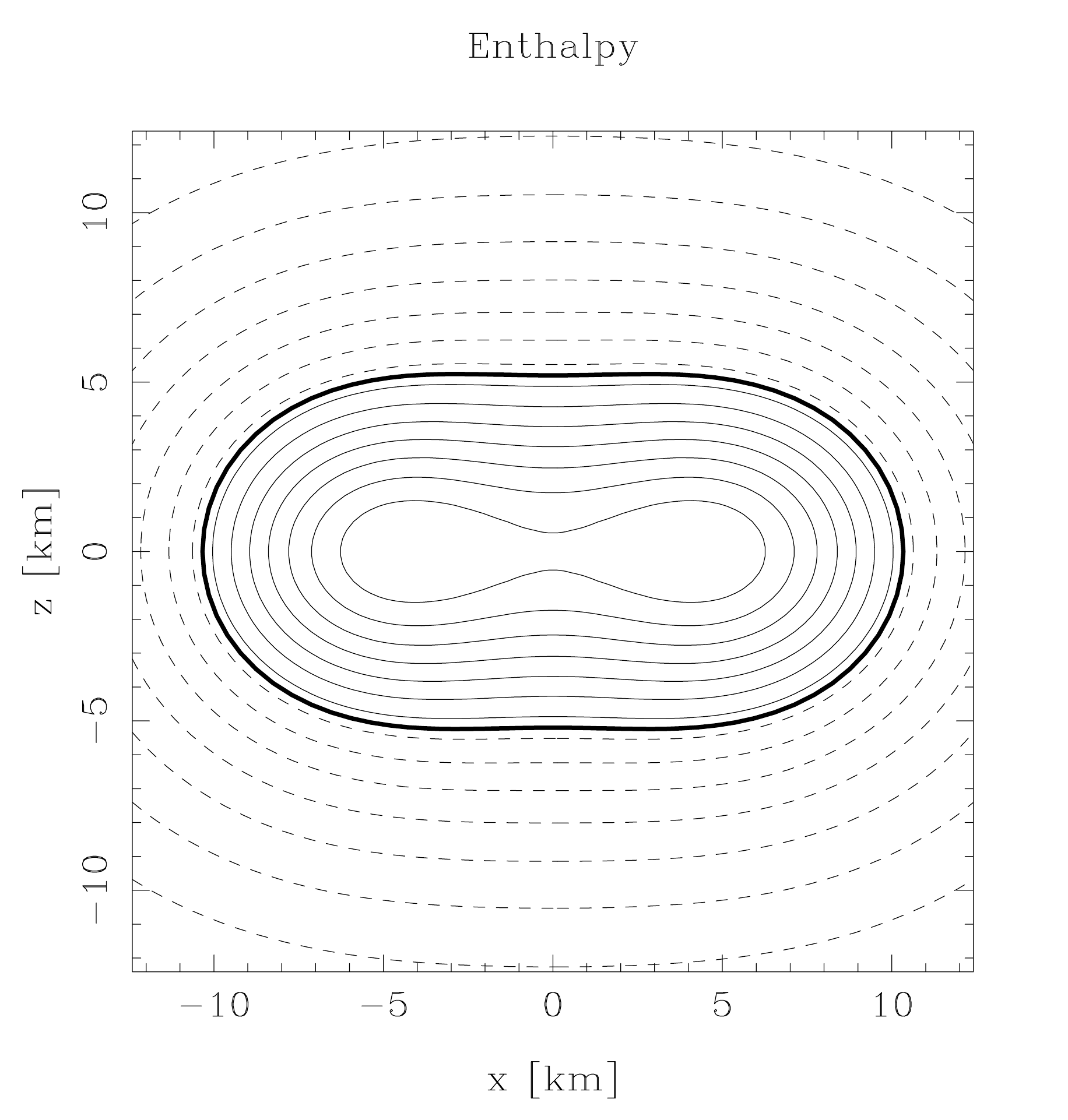}
\caption{\label{label} Baryon chemical potential isocontour lines for the most massive star reproduced by $\mu=3.5\times10^{32}$ A m$^2$.  \\ \\}
\end{minipage}
\hspace{2pc}%
\begin{minipage}{20pc}
\includegraphics[width=20.pc]{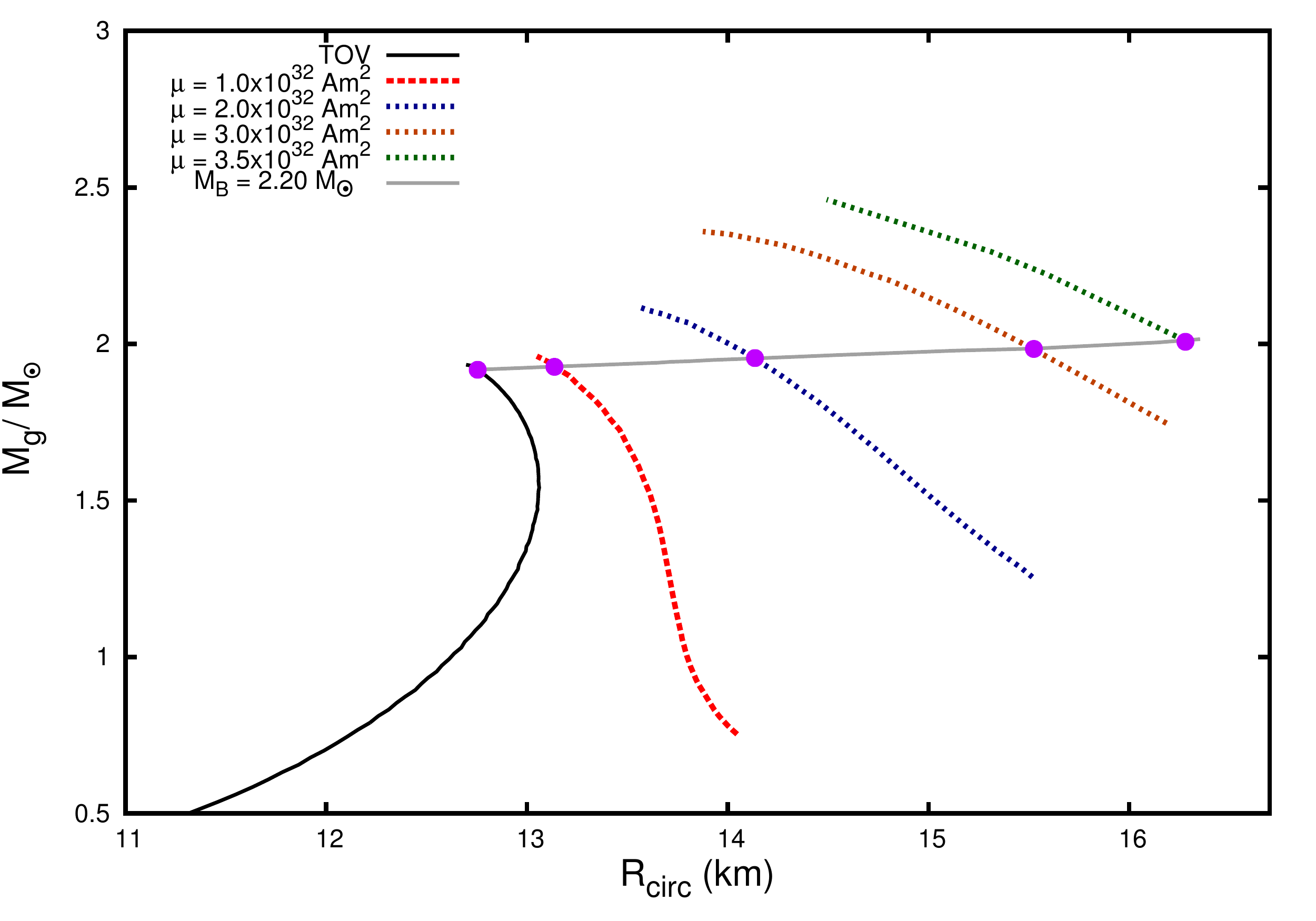}
\caption{\label{label} Mass-radius diagram for non-magnetized and magnetized models (with fixed magnetic moments). The gray line shows an equilibrium sequence for a fixed baryon mass of 2.2 M$_\odot$.}
\end{minipage} 
\end{figure}

To include the CMF EoS in the LORENE code, we calculate the EoS as a function of baryon chemical potential and magnetic field, as explained in Refs.~ \cite{Dexheimer:2011pz, Chatterjee:2014qsa, Franzon:2015sya}. Then, we solve the coupled Einstein-Maxwell's equations and determine the magnetic field in each part of a star. As a result, larger dipole magnetic moments $\mu$ (reproducing larger magnetic fields throughout the star) significantly increase the mass of the star. Using the maximum $\mu$ value allowed by the code, $\mu=3.5\times10^{32}$ A m$^2$, a maximum magnetic field of $\sim 2\times10^{18}$ G is found  in the stellar center. This configuration has a maximum stellar mass of about $0.6$ M$_\odot$ larger than the field-free case. Note that the effects solely due to magnetic field effects from the equation of state on the star mass are mainly due to the magnetization and are relatively small, being visible only for $\mu\gtrsim3\times10^{32}$ A m$^2$. Fig.~2 shows the deformation of the most massive star reproduced by $\mu=3.5\times10^{32}$ A m$^2$. In this case, the ratio between the magnetic pressure and the matter pressure in the center for this star is 0.79.

When we fix the baryon mass in order to reproduce the time evolution of an isolated star, the magnitude of magnetic field effects decreases considerably. In this case, a maximum magnetic field of $\sim 1\times10^{18}$ G in the stellar center is reproduced by using $\mu\gtrsim3.5\times10^{32}$ A m$^2$ and a fixed baryon mass of 2.2 M$_\odot$.  This configuration has a maximum stellar mass of about $0.1$ M$_\odot$ larger than the field-free case. Nevertheless, the radius of stars with fixed baryon mass is still very different for different magnetic field strengths, as can be seen in Fig.~3. The circunferencial radius is a coordinate-independent characterization of the stellar equator $C/(2\pi)$, with $C$ being the proper length of the circumference of the star in the equatorial plane.

\section{Stellar Population}
\hfill \break

Strong magnetic fields have a very large impact on the population of neutron stars. In addition to the effects already discussed in Section 2 related to the EoS, there are magnetic field effects related to the star structure. First, the central density (and respective central baryon chemical potential) in magnetized stars is substantially reduced. This can be seen in Fig.~4, where the different panels show that the population is less exotic (less hyperons or quarks) for larger dipole magnetic moments. This induces a much lower amount of strangeness in the center of the star. Looking at this in another way, exotic degrees of freedom appear over time (and in a larger portion of the star) as the stellar magnetic field decays.

\begin{figure}[t!]
\centering
\includegraphics[trim={0 1.75cm 0 0},clip,width=35pc]{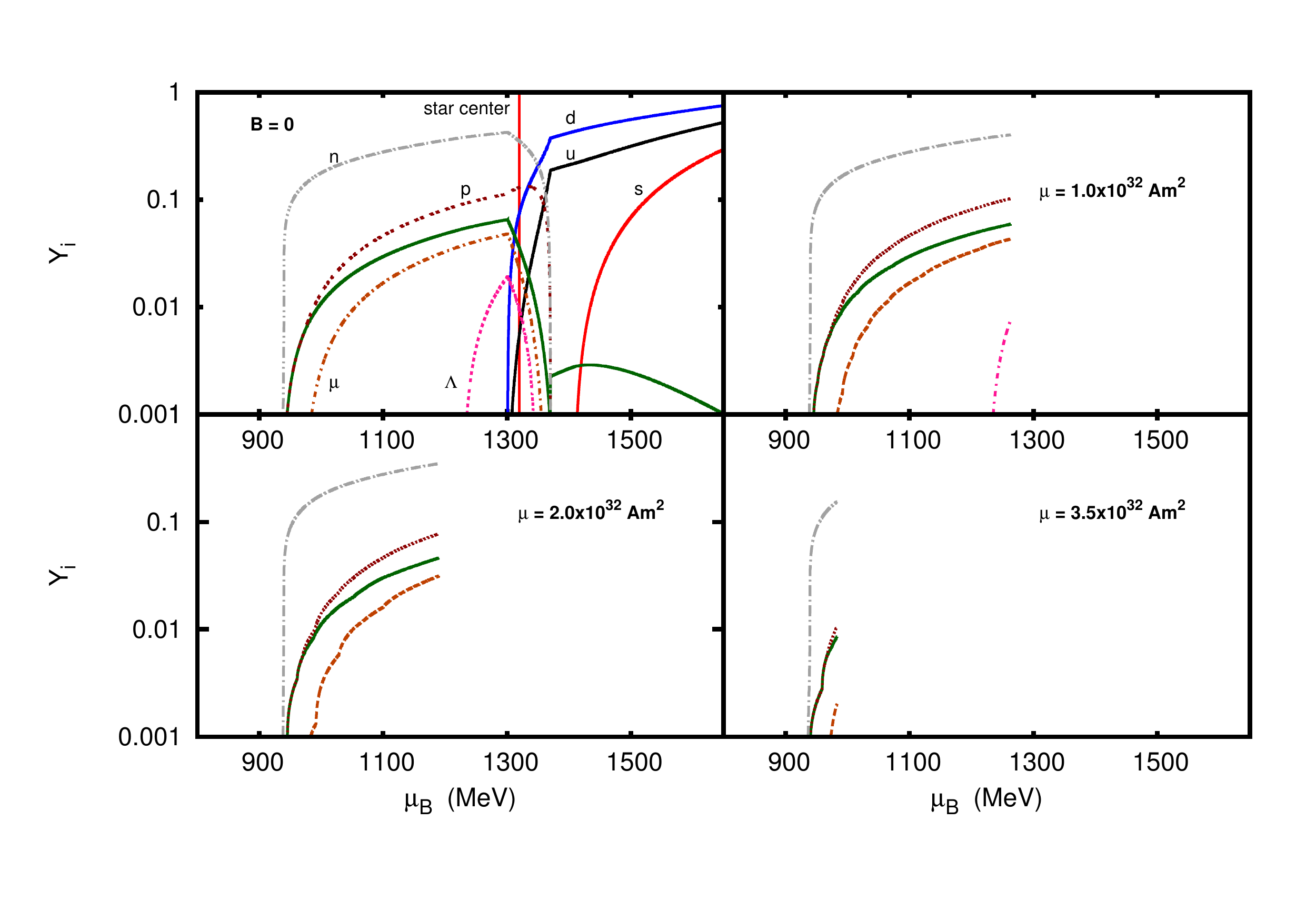}
\hspace{2pc}
\caption{\label{label} Stellar particle population as a function of the baryon chemical potential. Each panel shows a star with a different dipole magnetic moment. All figures represent an equilibrium sequence at fixed baryon mass $M_B=2.2$ M$_\odot$. For the non-magnetized case (B=0), the vertical red curve represents the chemical potential reached at the center of the star.}
\end{figure}

The second effect due to strong magnetic fields that is related to the star structure has to do with deformation. Quantities like the amount of trapped neutrinos and temperature, both associated with the cooling of stars, are quite different in deformed stars \cite{Franzon:2016iai}. To study these, we look at hot stars with fixed entropy per baryon $S/A=s/n_B=2$ and trapped neutrinos $Y_L=(n_e+n_{\nu_e})/n_B=0.4$, corresponding to the first evolutionary stage of neutron stars. Fig.~5 shows how the neutrinos are distributed along the equatorial ($\theta=\pi/2$) and polar ($\theta=0$) directions in the star for different current functions. At the center of the star, it can easily be noted that the amount of neutrinos is reduced by higher magnetic field distributions. Towards the outside part of the star,  the situation is more complicated. A higher magnetic field distribution implies a larger amount of neutrinos at a certain equatorial radius. This is related to the fact that the star becomes more oblate for larger $f_0$ and that the Lorenz force reverses its direction in the stellar equatorial plane, acting, therefore, differently at different radii.

Fig.~6 shows the temperature distribution in the star. Again, higher magnetic field distribution means a lower temperature at the center of the star, but a larger temperature at a larger equatorial radius of the stellar core. For a discussion on the relation of strong magnetic fields and star surface
thermal distribution see, for example, Ref.~\cite{Aguilera:2008ec}. The knowledge of the correct temperature distribution in proto-neutron and neutron stars is crucial for modeling the cooling of these stars. Thus, models that include the presence of strong magnetic fields
should be reconsidered, not only to investigate the effects of the anisotropy of the energy-momentum tensor due to the magnetic field, but also to include the asymmetric temperature
distribution in these objects.

\begin{figure}[t!]
\begin{minipage}{18pc}
\centering
\includegraphics[width=18.5pc]{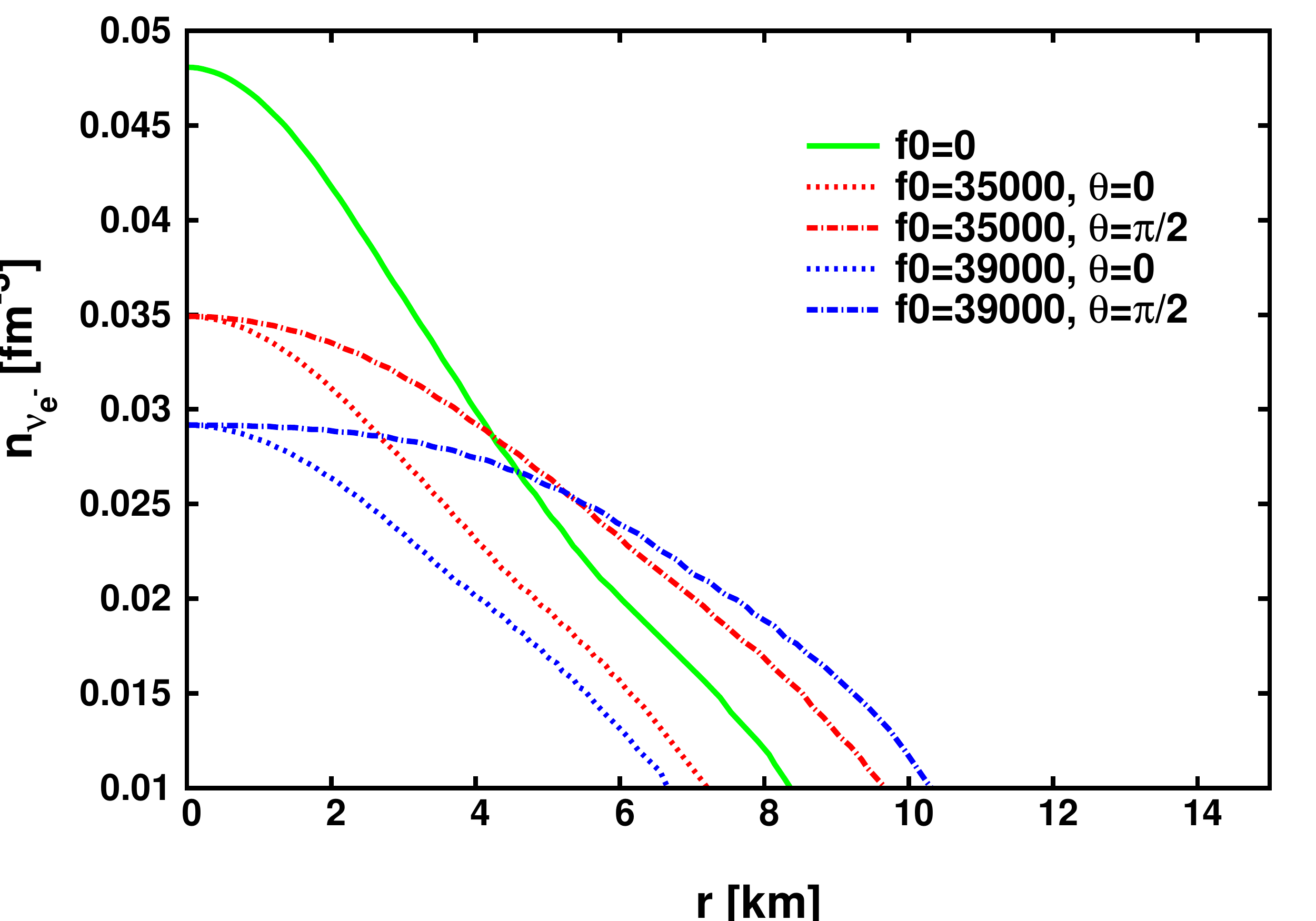}
\caption{\label{label} Electron neutrino density profile as a function of the equatorial and polar coordinate radii for one proto-neutron star with $M_B=2.35$ M$_\odot$ shown for different current functions.}
\end{minipage}
\hspace{2pc}%
\begin{minipage}{18pc}
\includegraphics[width=18.5pc]{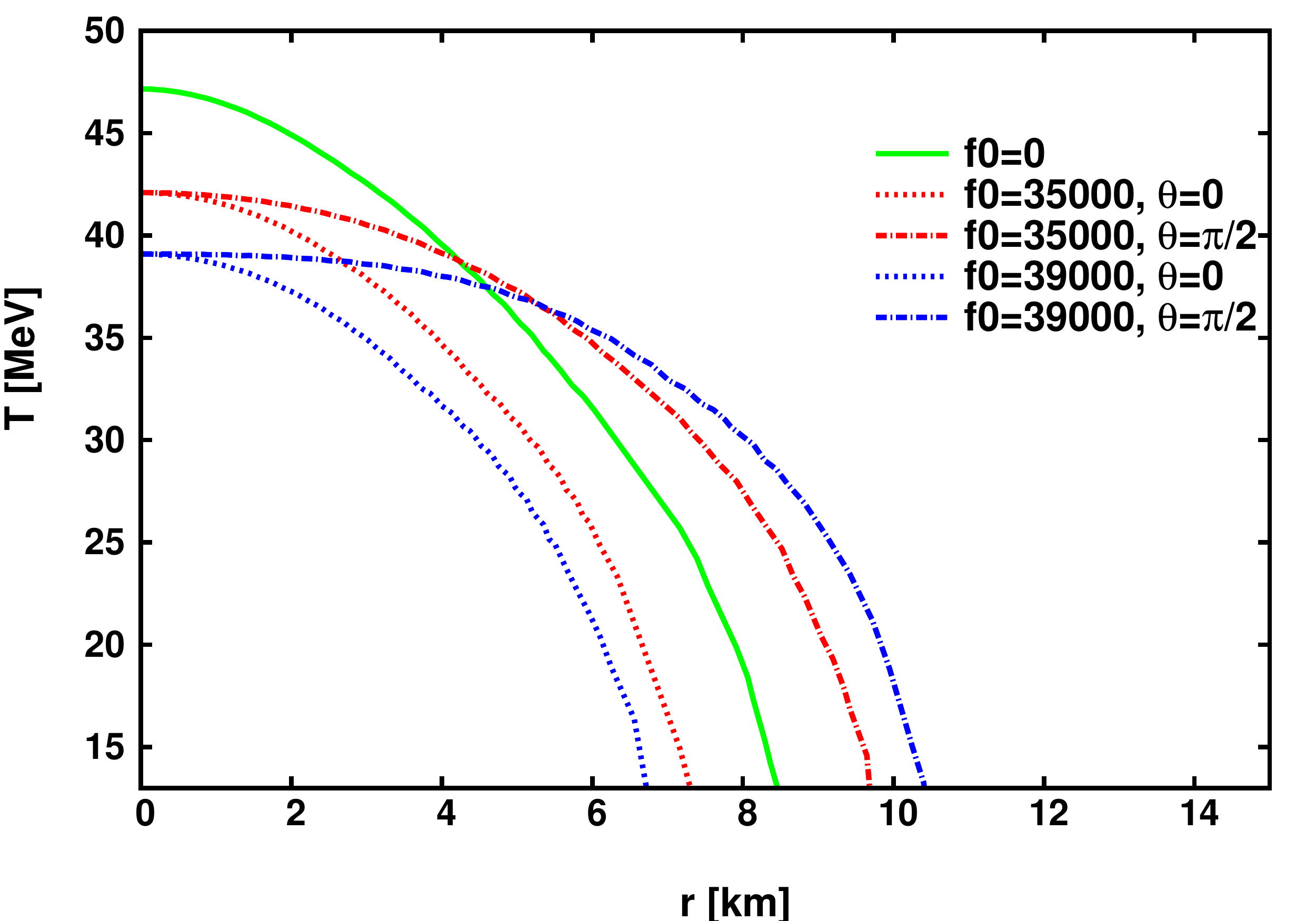}
\caption{\label{label} Temperature profile as a function of the equatorial and polar coordinate radii for one proto-neutron star with $M_B=2.35$ M$_\odot$ shown for different current functions.\\}
\end{minipage} 
\end{figure}

\section{Conclusions}
\hfill \break

Stars that possess strong magnetic fields need to be modeled very carefully. This includes direct effects from the magnetic field on the equation of state and also on the star structure. For example, we showed that stars that have a central magnetic field $\sim10^{18}$ G are already completely deformed and cannot be described through simple spherical solutions of Einstein's equations \cite{Oppenheimer:1939ne,Tolman:1939jz}. In this case, their population must be calculated using the magnetic field values and central density obtained from the axisymmetric solution of Einstein-Maxwell's equations. In our case, we model the microscopic EoS by using the CMF Model, which is an extended version of the non-linear realization of the SU(3) sigma model that also contains quarks.

Although the increase of stellar mass due to the magnetic field is small for a fixed baryon mass, the polar and equatorial radii are significantly modified, together with the amount of neutrinos and temperature profile, which now have to be calculated separately in different directions in the star. When we further assume that the stellar magnetic field decays over time throughout the star, we find that the appearance of exotic matter, such as hyperons, and phase transitions to quark matter take place at later evolution stages, when the central stellar density increases.

\section*{References}
\hfill \break

\bibliographystyle{iopart-num}
\bibliography{apssamp}

\end{document}